\documentstyle[emulateapj]{article}

\slugcomment{accepted for publication in ApJL}

\begin{document}
\title{The Ultraviolet flash accompanying GRBs from neutron-rich internal shocks}

\author{Y. Z. Fan$^{1,2,3}$ and D. M. Wei$^{1,2}$}
\affil{$^1$Purple Mountain Observatory, Chinese Academy of
Science, Nanjing 210008, China\\
       $^2$National Astronomical
Observatories, Chinese Academy of Sciences, Beijing, 100012,
China\\
$^3$On leave at Dept. of Physics, University of Nevada, Las Vegas, NV89154, USA \\
{yzfan@pmo.ac.cn~(yzfan@physics.unlv.edu)~~~~dmwei@pmo.ac.cn}}

\begin{abstract}
In the neutron-rich internal shocks model for Gamma-ray Burts
(GRBs), the Lorentz factors (LFs) of ions shells are variable, so
are the LFs of accompanying neutron shells. For slow neutron
shells with a typical LF $\sim {\rm tens}$, the typical
$\beta-$decay radius reads $R_{\rm \beta,s}\sim {\rm
several}\times10^{14}{\rm cm}$, which is much larger than the
typical internal shocks radius $\sim 10^{13}{\rm cm}$, so their
impact on the internal shocks may be unimportant. However, as GRBs
last long enough ($T_{90}>20(1+z){\rm s}$), one earlier but slower
ejected neutron shell will be swept successively by later
ejected ion shells in the range  $\sim 10^{13}{\rm cm}-10^{15}{\rm
cm}$, where slow neutrons have decayed significantly. We show
in this work that ion shells interacting with the $\beta-$decay
products of slow neutron shells can power a ultraviolet (UV) flash
bright to 12th magnitude during the prompt $\gamma-$ray emission
phase or slightly delayed, which can be detected by 
the upcoming Satellite SWIFT in the near future.

\end{abstract}


\section{Introduction}
As realized by many authors, the fireball of Gamma-ray bursts
(GRBs) may contain significant neutron component in essentially
all progenitor scenarios (e.g., Derishev, Kocharovsky $\&$
Kocharovsky 1999a, b; Beloborodov 2003a). The dynamics as well as
the observable signatures (the neutrino emission from the
proton-neutron decoupling and the changes in GRB lightcurves due
to neutrons preceding the ion shell or catching up it) of a
relativistic neutron-rich fireball have been investigated by
Derishev, Kocharovsky $\&$ Kocharovsky (1999a, b) firstly. Since
then, many authors have had their attention on the neutron-fed
GRBs. If the neutron
abundance is comparable to that of proton, the inelastic
collision between differentially streaming protons and neutrons in
the fireball will provide us observable 5-10 GeV neutrinos (Bachall \& M\'{e}sz\'{a}ros 2000, hereafter BM00; M\'{e}sz\'{a}ros \& Rees 2000).
The
baryon-loading problem in GRB models can be ameliorated if 
significant fraction of baryons confined in the
fireball are converted to neutrons (Fuller, Pruet $\&$ Abazajian 2000). Further investigations on the
observation implications have been presented in Pruet $\&$ Dalal (2002, hereafter
PD02) and Beloborodov (2003b, hereafter B03). In PD02, the short
GRBs are assumed to be powered by external shocks and the
neutron component lags behind the ion one. As the ion ejecta has
been decelerated significantly by the external medium, the decayed
neutron ejecta catches up the decelerated ion ejecta and powers
keV emission. In B03, the neutron-ejecta keeps ahead the ion-ejecta. 
The $\beta-$decay ($n\rightarrow p+e^{-}+\nu_{\rm
e}$) products share their momentum with the external medium immediately. As
a result, the external medium have been accelerated to a
ultra-relativistic velocity. The interaction between the ion ejecta and the accelerated medium is much different from the usual case (i.e., a fireball ejecta interacting with the static medium). So the presence of neutron ejecta
qualitatively changes the early dynamical evolution of GRBs'
remnant.

In PD02 and B03, only the ``fast'' neutron component has been
taken into account. In fact, the long, complex GRBs are more
likely to be powered by the interaction of shells with variant
LFs, i.e., the internal shocks model (Paczy\'{n}ski \& Xu 1994;
Rees \& M\'{e}sz\'{a}ros 1994). The best fit to the
multi-wavelength afterglows implies that a significant fraction of
the initial kinetic energy has been converted into internal energy
(Panaitescu $\&$ Kumar 2002), which requires that the difference
in velocities between the shells is significant (i.e., the
corresponding LFs satisfy $\eta_{\rm f}\gg \eta_{\rm s}$) and
their masses are comparable ($m_{\rm f}\approx m_{\rm s}$) (Piran
1999, hereafter P99) [Throughout this work, the subscript ${\rm
f,~s}$ represent the fast/slow shells respectively; ${\rm n,~p}$
represent the neutron/proton component respectively]. Typically,
$\eta_{\rm s}$ is in order of tens, and $\eta_{\rm f}$ is in order
of hundreds. Thus, for the slow shell, the n, p component coast
with $\Gamma_{\rm n,s}=\Gamma_{\rm p,s}\simeq\eta_{\rm s}$ since
$\eta_{\rm s}<\eta_\pi\simeq 3.9\times 10^2L_{52}
^{1/4}r_{0,7}^{-1/4}[(1+\xi)/2]^{-1/4}$ (BM00), where $L$ being
the total luminosity of the ejecta, $r_0$ being the radius of the
central engine, $\xi$ being the ratio of the neutrons to the
protons contained in the shells. In this Letter, we adopt the
convention $Q_{\rm x}=Q/10^{\rm x}$ for expressing the physical
parameters, using cgs units. For the fast shell, the n, p
components move with different LFs: $\Gamma_{\rm n,f}<\Gamma_{\rm
p,f}$ since generally $\eta_{\rm f}>\eta_{\rm \pi}$ (BM00). In the
internal shocks phase, the fast ion shell catches the slower but
earlier one at a radius $R_{\rm int}\sim 10^{13}{\rm cm}$ and forms
a new one moving with a LF $\Gamma_{\rm m}\sim {\rm
several~hundred}$ (hereafter the formed ``new'' shell is named as
the ``i-shell''). At $R_{\rm int}$, the $\beta-$decay of the
neutron component is far from important unless the typical
variability timescale is longer than 0.1 s (See Rossi, Beloborodov
\& Rees 2004 for more information).

As the duration of GRB is long enough, the much later
ejected i-shell catches up the earlier slow neutron shell
(hereafter, the ``n-shell'') at $R_{\rm cat}\approx 2\Gamma_{\rm
n,s}^2c \delta T/(1+z)$, where $\delta T$ being the ejection
time-lag of the earlier slow n-shell and the later i-shell; On the
other hand, the $\beta-$decay radius of slow neutrons
reads $R_{\rm \beta,s}\approx 2.6\times 10^{13}\Gamma_{\rm
n,s}~{\rm cm}$, at which slow neutrons have decayed significantly.
As long as $R_{\rm cat}$ is comparable with $R_{\rm \beta,s}$,
i.e., $\delta T\geq 14(1+z){\Gamma^{-1}_{\rm n,s,1.5}}$, the 
decayed products of earlier but slower ejected neutron  shells will
be swept orderly by the later ejected ion shells in a range
$\simeq R_{\rm int}-R_{\rm \beta,s}$. The possible emission
signature powered by that interaction is of our interest in this
Letter. Otherwise, for GRBs much shorter than
$14(1+z){\Gamma^{-1}_{\rm n,s,1.5}}{\rm s}$, the last i-shell
crosses the first slow n-shell at a radius $\ll R_{\rm \beta,s}$,
where the $\beta-$decay is unimportant and there is no 
observable signature.

\section{The simplest model}
Even for a neutron-free internal shocks model, the practical LF
distribution of shells involved is far from clear so far, let
alone say the neutron-rich one. Here we use an approximate model
proposed by Guetta, Spada $\&$ Waxman (2001), in which the LFs of
i-shells are drawn from a bimodal distribution, $\Gamma_{\rm
ej}=\eta_{\rm f}$ or $\Gamma_{\rm ej}=\eta_{\rm s}$, with equal
probability. That simple model is favored by its relative high
energy conversion efficiency and peak energy. In this Letter,
mainly for simplicity, we make the following assumptions: (i)
$\xi=1$, and the ratios between the mass of the fast neutrons,
slow neutrons and i-shells are 1:1:2. (ii) Each fast (slow)
neutron/ion shell moves with a same LF $\Gamma_{\rm n,f}\sim 300$
($\Gamma_{\rm n,s}\sim 30$) and $\Gamma_{\rm p,f}\sim 1000$
($\Gamma_{\rm p,s}\sim 30$) respectively. After the merger of a
pair of fast/slow ion shells, the resulting i-shell moves with
$\Gamma_{\rm m}\sim 200$. (iii) The energetic ejecta are expanding
into the low density interstellar medium (ISM), which is the case
for most GRBs (Panaitescu $\&$ Kumar 2002). (iv) The duration of
the GRB is longer than $30(1+z){\rm s}$ since the constraint
$\delta T\simeq 14(1+z) {\Gamma_{\rm n,s}^{-1}}_{1.5}~{\rm s}$
~should be satisfied.

The $\beta-$decay radius of fast neutrons $R_{\rm \beta,f}\approx
8\times 10^{15}{{\Gamma_{\rm n,f}}_{,2.5}}~{\rm cm}$ is much
larger than $R_{\rm \beta,s}$. We ignore them since for $R\leq
R_{\rm \beta,s}$, there is only a small fraction of fast neutrons
decayed. The ISM has been swept by the early and fast ions as well
as the decayed products of fast neutrons, so the decay products of
slow neutrons move freely.

Assuming that at radius $R$, the $(k-1)$th i-shell crosses the $j$th 
slow n-shell ($j,~k\gg1$). At radius
$\simeq R+\Delta R$, the $k$th i-shell will catch up the $j$th
slow n-shell, where $\Delta R\approx 2\Gamma_{\rm n,s}^2\Delta$,
$\Delta\approx c\delta t/(1+z)$, $\delta t\sim 0.01(1+z){\rm s}$
being the typical variability timescale of GRBs. For the $j$th
n-shell, the products of slow neutrons decayed at radius $\leq R$
have been carried away by i-shells ejected early than the $k$th
one. Hence the mass swept by $k$th i-shell is $\Delta M\simeq
{M_{\rm n}\Delta R\over R_{\rm \beta,s}}$, where $M_{\rm n}\equiv
M_{\rm n}^0\exp(-{R\over R_{\rm \beta,s}})$, $M_{\rm n}^0$ being
the the initial rest mass of the n-shell. So the averaged comoving
number density of proton (due to $\beta-$decay) swept by $k$th
i-shell can be estimated by
\begin{equation}
n\approx {\Delta M\over 4\pi R^2m_{\rm p} \Gamma_{\rm n,s}\Delta}
\approx {\Gamma_{\rm n,s}M_{\rm n}\over 2\pi R^2m_{\rm p}R_{\rm
\beta,s}}.\label{Eq-n}
\end{equation}

Notice that in front of the much later ejected i-shells, there are
hundreds of decaying n-shells. With assumptions made before, the
whole process of each i-shell interacting with these decaying
n-shells is rather similar. For convenience, in our following
treatment, the discrete interaction of each i-shell with these
decaying n-shells has been simplified as an i-shell sweeping a
moving proton trail (the LF of which is $\Gamma_{\rm n,s}$) with a
number density $n$ continually. The dynamics of that interaction
can be described by the shock model as follows: Now
the trail is moving with a LF $\Gamma_{\rm n,s}\sim {\rm tens}$,
the generated thermal energy (comoving frame) in the shock
front satisfies $dU=(\gamma_{\rm rel}-1)dm~c^2$ (e.g., B03) rather
than $dU=(\gamma-1)dm~c^2$; where $\gamma_{\rm
rel}=\gamma\Gamma_{\rm n,s}(1-\beta_{\rm \gamma}\beta_{\rm
\Gamma_{\rm n,s}})$ being the LF of the decelerating i-shell (the
LF of which is $\gamma$) relative to the trail; $\beta_{\rm
\gamma}$ and $\beta_{\rm \Gamma_{\rm n,s}}$ being the
corresponding velocity of $\gamma$ and $\Gamma_{\rm n,s}$; the
mass swept by the i-shell reads
\begin{equation}
dm=4\pi n m_{\rm p} \Gamma_{\rm n,s} R^2
(\beta_{\gamma}-\beta_{\Gamma_{\rm n,s}})dR\simeq {M_{\rm n}\over
R_{\rm \beta,s}}dR.
\end{equation}
After some simple algebra, the energy conservation of the system (the
decelerating i-shell and the swept slow neutron trail) yields
\begin{equation}
{d\gamma \over dm}=-{{\gamma\gamma_{\rm rel}-\Gamma_{\rm
n,s}}\over M_{\rm ion}+m+(1-\epsilon)U}, \label{Dynamics}
\end{equation}
where $M_{\rm ion}$ being the rest mass of the i-shell; $\epsilon$
being the radiation efficiency of the shock. For $\Gamma_{\rm
n,s}=1$, equation (\ref{Dynamics}) reduces to the familiar form
${d\gamma \over dm}=-{{\gamma^2-1}\over M_{\rm
ion}+m+(1-\epsilon)U}$ (P99).

In the downstream, the electrons have been heated by the shock. As
usual (e.g., P99), we assume that the shocked electrons distribute
as ${dn\over d\gamma_{\rm e}}\propto \gamma_{\rm e}^{-p'}$ for
$\gamma_{\rm e}>\gamma_{\rm e,m}$, where $p'\sim 2.3$ being the
typical power law index, $\gamma_{\rm e,m}\approx \epsilon_{\rm
e}{p'-2\over p'-1}{m_{\rm p}\over m_{\rm e}}(\gamma_{\rm rel}-1)$;
$\epsilon_{\rm e}$ and $\epsilon_{\rm B}$ being the fractions of
the shock energy given to  electrons and the random magnetic field at shock,
respectively. The
comoving downstream magnetic field $B\approx \sqrt{32\pi
\epsilon_{\rm B}\gamma_{\rm rel}(\gamma_{\rm rel}-1)nm_{\rm
p}c^2}$. As shown in P99, there is a critical LF, above which
synchrotron radiation is significant: $\gamma_{\rm e,c}={6\pi
m_{\rm e}c\over (1+Y)\sigma_{\rm T }\gamma B^2t}$, where $t=t_{\rm
obs}/(1+z)$ is determined by $dR=2\gamma^2 cdt$ ($t_{\rm obs}$
being the observer time, $z\sim1$ being the red-shift of GRBs),
$\sigma_{\rm T}$ being the Thomson cross section, $Y$ being the
Compton parameter (e.g., Wei \& Lu 1998, 2000; Sari \& Esin 2001):
$Y\simeq {-1+\sqrt{1+4x\epsilon_{\rm e}/\epsilon_{\rm B }}\over
2}$, where $x$ being the radiation coefficient of electrons, so
$\epsilon\equiv x\epsilon_{\rm e}$. For $\gamma_{\rm
e,m}>\gamma_{\rm e,c}$ (fast cooling), $x=1$; For $\gamma_{\rm
e,m}<\gamma_{\rm e,c}$, $x={(\gamma_{\rm e,m}/\gamma_{\rm
e,c}})^{p'-2\over 2}$ (slow cooling).

\begin{figure}
\begin{picture}(100,450)
\put(0,0){\includegraphics{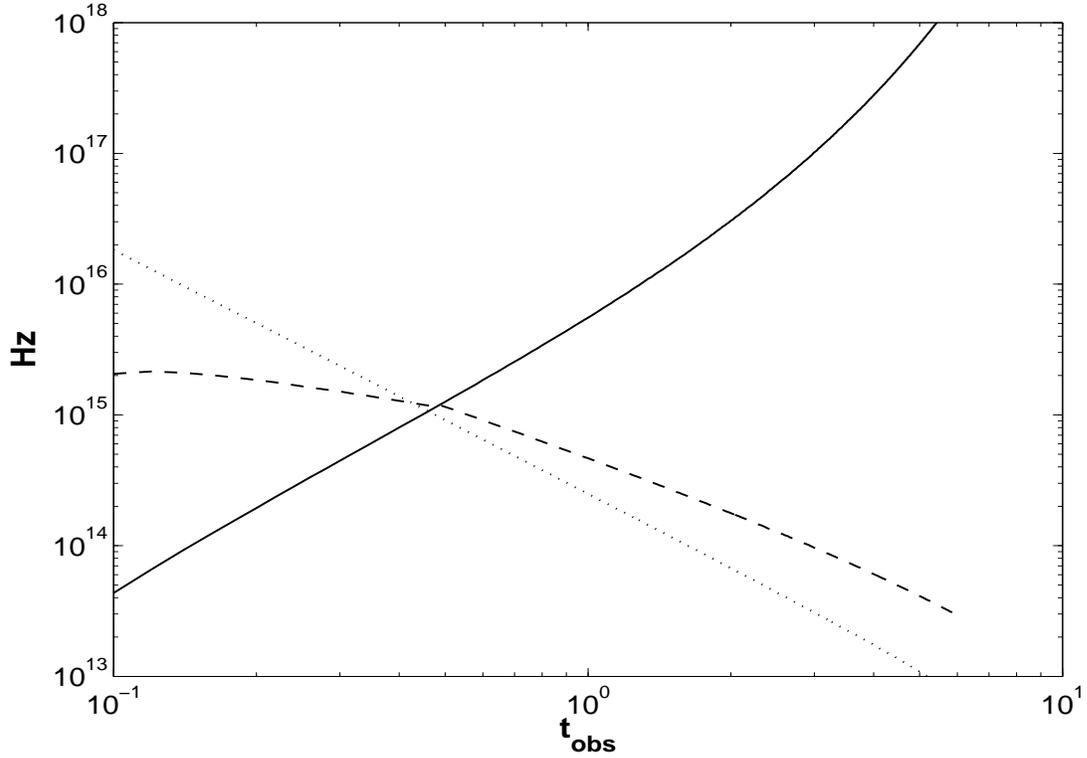}}
\end{picture}
\caption{The radiation frequencies involved in an i-shell interacting with the trail of slow n-shells as a function of
observer time. The physical parameters are taken as: $M_{\rm ion}=2M_{\rm
n}^0=5.6\times10^{26}{\rm g}$, which is corresponding to a
Luminosity $L=10^{52}{\rm ergs~s^{-1}}$ and $\delta
t=10^{-2}(1+z){\rm s}$, $\Gamma_{\rm m}=200$; $\Gamma_{\rm
n,s}=30$. $\epsilon_{\rm e}=0.3$, $\epsilon_{\rm B}=0.01$, $z=1$ ($D_{\rm L}=2.2\times 10^{28}{\rm cm}$).
The solid line, dashed line and dotted line represent $\nu_{\rm c}$ (the cooling frequency),
$\nu_{\rm a}$ (the synchrotron self-absorption frequency) and $\nu_{\rm m}$ (the typical synchrotron 
frequency), respectively. 
 The start point has been chosen to be $R=10^{14}$cm,
the corresponding start time $t_{\rm obs,beg}=0.04(1+z){\rm s}$.
The end point has been chosen as $2.5\times10^{15}{\rm cm}\simeq
3R_{\rm \beta,s}$, the corresponding end time $t_{\rm
obs,end}\simeq 3(1+z){\rm s}$.} \label{Frequency}
\end{figure}

Given a proper boundary
condition, equation (\ref{Dynamics}) can be easily solved. 
With resulting $\gamma_{\rm e,m}$, $\gamma_{\rm
e,c}$ and $B$, we can calculate the typical synchrotron radiation
frequency $\nu_{\rm m}={\gamma_{\rm e,m}^2\gamma eB\over 2(1+z)\pi
m_{\rm e}c}$ ($e$ being the charge of electron), the cooling
frequency $\nu_{\rm c}={\gamma_{\rm e,c}^2\gamma eB\over 2(1+z)\pi
m_{\rm e}c}$ and the self-absorption frequency $\nu_{\rm a}$ (the
detailed calculation on $\nu_{\rm a}$ can be found in the appendix
of Wu et al. (2003)), with which we can analytically calculate the
synchrotron radiation flux at the fixed band. The numerical results 
for these typical frequencies have been plotted in figure \ref{Frequency}, 
where the parameters are taken as: $M_{\rm
ion}=2M_{\rm n}^0=5.6\times10^{26}{\rm g}$, which is corresponding
to a Luminosity $L=10^{52}{\rm ergs~s^{-1}}$ and $\delta
t=10^{-2}(1+z){\rm s}$, $\Gamma_{\rm m}=200$; $\Gamma_{\rm
n,s}=30$, $\epsilon_{\rm e}=0.3$, $\epsilon_{\rm B}=0.01$, $z=1$
($D_{\rm L}=2.2\times 10^{28}{\rm cm}$). 
 As shown in figure \ref{Frequency}, at the early time, the electrons are usually in fast
cooling case and $\nu_{\rm a}$ is above the optical band. Lately,
the electrons are in slow cooling case and $\nu_{\rm a}$ drops to
$\sim {\rm a~few}\times 10^{13}{\rm Hz}$ (Currently, the observed
emission are coming from hundreds of i-shells interacting with the
slow neutrons trail. For the same reason, in our estimating the
self-absorption frequency, the total number of electrons
contributed has been assumed to be $300$ times that of one trail.
In fact, if just an i-shell interacting with the slow neutron
trail has been taken into account, $\nu_{\rm a}$ should be one
order or more less than the value presented here).

The synchrotron flux as a function of observer frequency can be
approximated as follows: In the case of fast cooling (Just for $\nu_{\rm a}>\min\{\nu_{\rm c},~\nu_{\rm m}\}$. For $\nu_{\rm a}<\min\{\nu_{\rm c}, \nu_{\rm m}\}$, please refer to Eqs. (4-5) of Zhang \& M\'{e}sz\'{a}ros (2004)),
\begin{equation}
 F_{\rm \nu}\approx{F_{\rm max}} \left \{
   \begin{array}{llll}
   {({\nu_{\rm c}/\nu_{\rm
a}})^{3}(\nu/\nu_{\rm c})^{2}}, \,\,\,\, & (\nu<\nu_{\rm c}),\\
   {({\nu_{\rm a}/\nu_{\rm
c}})^{-1/2}(\nu/\nu_{\rm a})^{5/2}}, \,\,\,\, & (\nu_{\rm
c}<\nu<\nu_{\rm a}),\\
({\nu/\nu_{\rm
c}})^{-1/2},\, \,\,\,\, & (\nu_{\rm a}<\nu<\nu_{\rm m}), \\
 ({\nu_{\rm m}/\nu_{\rm
c}})^{-1/2}({\nu/\nu_{\rm m}})^{-p'/2},\,\,\,\,\, & (\nu>\nu_{\rm
m}).
   \end{array}
   \right.
\end{equation}

In the case of slow cooling
\begin{equation}
 F_{\rm \nu}\approx{F_{\rm max}} \left \{
   \begin{array}{llll}
   ({\nu_{\rm m}\over \nu_{\rm a}})^{\rm
{(p'+4)\over 2}}({\nu\over \nu_{\rm m}})^{2}, \,\,\,\, &
(\nu<\nu_{\rm
m}),\\
   ({\nu_{\rm a}\over \nu_{\rm m}})^{-\rm
{(p'-1)\over 2}}({\nu\over \nu_{\rm a}})^{5/2}, \,\,\,\, &
(\nu_{\rm m}<\nu<\nu_{\rm
a}),\\
({\nu\over \nu_{\rm
m}})^{\rm -{(p'-1)\over 2}},\, \,\,\,\, & (\nu_{\rm a}<\nu<\nu_{\rm c}), \\
 ({\nu_{\rm c}\over \nu_{\rm m}})^{\rm
-{(p'-1)\over 2}}({\nu\over \nu_{\rm c}})^{\rm -{p'\over 2}},\,
\,\,\,\, & (\nu>\nu_{\rm c}).
   \end{array}
   \right.
\end{equation}

Approximately, $F_{\rm max}={3\sqrt{3}\Phi_{\rm p'}(1+z)N_{\rm e}
m_{\rm e}c^2\sigma_{\rm T}\over 32\pi^2 eD_{\rm L}^2}\gamma B$,
$\Phi_{\rm p'}$ being a function of $p'$ (for $p'\sim 2.3$,
$\Phi_{\rm p'}\simeq 0.6$); $N_{\rm e}$ being the number of
electrons involved in the emission, $D_{\rm L}$ being the
luminosity distance (Wijers \& Galama 1999).

\begin{figure}
\begin{picture}(100,350)
\put(0,0){\includegraphics{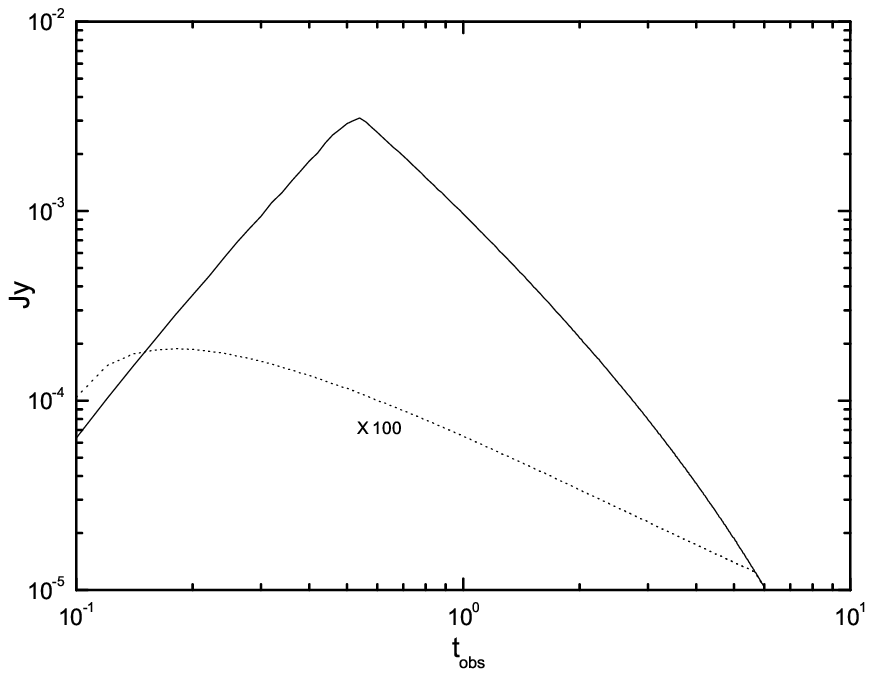}}
\end{picture}
\caption{The synchrotron radiation powered by an i-shell
interacting with the trail of slow n-shells as a function of
observer time. The parameters are  the same as those taken in Fig. \ref{Frequency}. 
 The solid line, dotted line
represent $\nu_{\rm obs}=10^{15}{\rm Hz}$ and 5 keV respectively.
For clarity, the value of dotted line has been multiplied by a
factor 100.  \textit{Please note that what we
observed is the emission powered by a series (several thousand) of
i-shells interacting with the slow neutrons' trail rather than
just by one. In the simplest model proposed here, about 300 sample
light curves overlap, each lags with $\delta t=10^{-2}(1+z){\rm
s}$. The resulting net observable fluxes at $10^{15}{\rm Hz}$ and
5keV are about $0.11{\rm Jy}$ and $1.4\times 10^{-9}{\rm
ergs}~{\rm cm^{-2}}~{\rm s^{-1}}$ respectively.} } \label{Lightcurve}
\end{figure}

The upcoming Satellite Swift carries three
telescopes\footnote{http://swift.gsfc.nasa.gov/docs/swift/}:
The Burst Alert Telescope (BAT), the X-ray Telescope (XRT), the
Ultraviolet and Optical Telescope (UVOT). The energy range of XRT
is 0.2-10 keV. UVOT covers 170nm-650nm with six colors. So in
Figure \ref{Lightcurve} we calculate the emission at the observer
frequency $\nu_{\rm obs,1}=1.0\times 10^{15}{\rm Hz}$ and
$\nu_{\rm obs,2}=5{\rm keV}$.

The synchrotron radiation of an i-shell interacting with the trail
as a function of the observer time has been shown in Figure
\ref{Lightcurve}. What we
observed is the emission comes from a series (several thousand) of
i-shells interacting with the trail of slow neutrons rather than
just from one. In the simplest model proposed here, $\sim {3{\rm
s}(1+z)/\delta t}=300$ sample light curves overlap, each with a
$\delta t/(1+z)=10^{-2}{\rm s}$ delay. The resulting net
observable fluxes at $10^{15}{\rm Hz}$ and 5keV are about
$0.11{\rm Jy}$ and $1.4\times 10^{-9}{\rm ergs}~{\rm cm^{-2}}{\rm
s^{-1}}$ respectively. The signature of 5keV emission is only
marginal since the typical X-ray component during many GRBs is
strong, too. So we do not discuss it any longer. The sensitivity
of UVOT is about 19th magnitude in a 10-s exposure. The
spacecraft's time-to-target is about 20$-$70 seconds, in
principle, part of, if not all the emission predicted here can be
detected in the near future. However, we need to investigate
whether some other emissions during prompt $\gamma-$ray
emission phase can dominate over the emissions predicted here or
not. Recently, it is suggested that the
relativistic $e^{\pm}$ pair generated in the internal shocks phase
can power a UV flash $\sim 13{\rm th}$ magnitude, too (Fan \& Wei 2004). But the
actual value of that emission is proportional to $\Gamma_{\rm
m}^4[\delta t/(1+z)]^{5/2}$, and in that work the fiducial value
for these parameters are taken as $300$ and $0.1(1+z){\rm s}$
respectively. If $\delta t \sim 0.01(1+z){\rm s}$ or smaller, as
taken in this work, the UV flash predicted in Fan \& Wei (2004)
will be much weaker ($\lesssim 18{\rm th}$ magnitude) unless
$\Gamma_{\rm m}$ is much larger than 300. Thus the UV emission
predicted in this work can be detected independently.

\section{Discussion}
In the standard baryonic internal shocks model for Gamma-ray
bursts (Paczy\'{n}ski \& Xu 1994; Rees \& M\'{e}sz\'{a}ros 1994),
the prompt $\gamma-$ray emission are powered by the interaction
between ion shells with variable LFs. If these shells are neutron
rich, the neutron component coupled with the ''fast ions'' is
accelerated to a larger bulk LF ($\sim {\rm several~hundred}$),
i.e., the fast neutron component, which is the focus of many
publications (e.g., PD02, B03). However, the neutron component
coupled with the ``slow ions'' can only be accelerated to a
moderate bulk LF $\sim {\rm tens}$, for which the typical
$\beta-$decay radius $\sim 10^{15}{\rm cm}$ is significantly
smaller than that of the fast neutrons. As long as the duration of
GRB is long enough, the earlier n-shells will be swept by the much
later i-shells successively till the radius $\simeq R_{\rm
\beta,s}$, where the slow neutrons have decayed significantly.
Consequently the interaction may power observable emission.

In this Letter, with an extremely simplified model to describe the
later but faster i-shells interacting with the much early but
slow, decaying n-shells, we have shown that there comes a
$\sim$12th magnitude UV flash at the later time of long GRBs. That
emission is bright enough to be detected by the upcoming Satellite
SWIFT in the near future. The emission predicted here is
independent on the poorly known medium distribution around the
progenitor of GRBs, as long as it is not very dense; it is also
independent of some other physical processes, such as the
$e^{\pm}$ pair emission (Fan \& Wei 2004), the possible reverse
shock emission (As usual, if we assume the number density of
interstellar medium (ISM) is about $1~{\rm cm}^{-3}$, $R_{\rm
\beta,s}$ is much smaller than the deceleration radius $\sim
10^{17}{\rm cm}$, at which about half of the kinetic energy of the
ejecta has been converted into the thermal energy of the shocked
ISM, for a neutron-free ejecta) and so on.

The resulting net light-curve in this work increases rapidly
firstly. Then there comes a flat lasting $\sim T_{90}$, the
duration of corresponding GRB. Finally, the light-curve drops
sharply, where the observed emission are mainly contributed by the
"Equal-arriving surface" effect (e.g., Kumar \& Panaitescu 2000;
Fan, Wei \& Wang 2004), if other emission such as the reverse
shock emission has not been taken into account. It should be noted
that the model taken here is a great simplification of the real
situation, in which both the ion/neutron shells are moving with
variable LFs, it is natural to expect that the corresponding
emission is variable with time, too.  However, to simulate such
complicate processes is far beyond the scope of this Letter.
Nonetheless, the actual flux may not be far dimmer than our
prediction since averagely speaking, the LFs of the slow n-shells
and the i-shells are significantly different, but their masses are
comparable. So the interaction of them can power a bright UV flash
in the range $\sim 10^{13}-10^{15}{\rm cm}$.

\acknowledgments Y. Z. F thanks T. Lu and Z. Li for their long-term
encouragement on the subject of neutron-fed GRBs. We also thank
the anonymous referee for her/his constructive comments. This work is
supported by the National Natural Science Foundation (grants
10225314 and 10233010), the National 973 Project on Fundamental
Researches of China (NKBRSF G19990754). This work is in memory of
 Changlin, Zhang (Chang'an University, China), Y. Z. F's university physics
teacher, who has passed away on 2004 September 6.

\begin{thebibliography}{99}
\bibitem[]{427} Bahcall, J. N., $\&$ M\'{e}sz\'{a}ros, P. 2000, Phys.
Rev. Lett., 85, 1362 (BM00)
\bibitem[]{429} Beloborodov, A. M. 2003a, ApJ, 588, 931
\bibitem[]{430}  ------. 2003b, ApJ, 585, L19 (B03)
\bibitem[]{431} Derishev, D. E., Kocharovsky, V. V., $\&$ Kocharovsky, ${\rm V_
L}$. V. 1999a, A$\&$A, 345, L51
\bibitem[]{433}  ------. 1999b, ApJ, 521, 640
\bibitem[]{434} Fan, Y. Z., \& Wei, D. M. 2004, MNRAS, 351, 292
\bibitem[]{435} Fan, Y. Z., Wei, D. M., \& Wang, C. F. 2004, A\&A, 424,
477
\bibitem[]{437} Fuller, G. M., Pruet, J., $\&$ Abazajian, K. 2000, Phys.
Rev. Lett., 85, 2673
\bibitem[]{439} Guetta, D., Spada, M., $\&$ Waxman, E. 2001, ApJ, 557, 399
\bibitem[]{440} Kumar, P., \& Panaitescu, A. 2000, ApJ, 541, L9
\bibitem[]{441} M\'{e}sz\'{a}ros, P., $\&$ Rees, M. J. 2000, ApJ,
541, L5
\bibitem[]{443} Paczy\'{n}ski, B.,  $\&$ Xu, G. H. 1994, ApJ, 427,
708
\bibitem[]{445} Panaitescu, A., $\&$ Kumar, P. 2002, ApJ, 571, 779
\bibitem[]{446} Piran, T. 1999, Phys. Rep., 314, 575 (P99)
\bibitem[]{447} Pruet, J., $\&$ Dalal, N. 2002, ApJ, 573, 770 (PD02)
\bibitem[]{448} Rees, M. J., \& M\'{e}sz\'{a}ros, P. 1994, ApJ, 430,
L93
\bibitem[]{450} Rossi, E. M.,  Beloborodov, A. M. \& Rees, M. J.
2004, to appear in the 2003 Santa Fe GRB Conference Proceedings
(astro-ph/0401355)
\bibitem[]{453} Sari, R., \& Esin, A. A. 2001, ApJ, 548, 787
\bibitem[]{454} Wei, D. M., \& Lu, T. 1998, ApJ, 505, 252
\bibitem[]{455}  ------. 2000, A\&A, 360, L13
\bibitem[]{456} Wijers, R. A. M. J., \& Galama, T. J. 1999, ApJ, 523, 177
\bibitem[]{457} Wu, X. F., Dai, Z. G., Huang, Y. F., \& Lu, T. 2003,
MNRAS, 342, 1131
\bibitem[]{459} Zhang, B., \& M\'{e}sz\'{a}ros, P. 2004, Int. J. Mod.
Phy. A., 19, 2385
\end{thebibliography}
\end{document}